\documentclass{iopart}
\usepackage{iopams}
\usepackage{setstack}
\usepackage{url}
\usepackage{graphicx}	

\begin{document}


\title[The interior structure of slowly rotating black holes]{The interior structure of slowly rotating black holes}

\author{Andrew J S Hamilton}
\address{JILA and
Dept.\ Astrophysical \& Planetary Sciences,
Box 440, U. Colorado, Boulder, CO 80309, USA}
\ead{Andrew.Hamilton@colorado.edu}

\newcommand{\simpropto}{
\begin{array}{c}
\propto \\[-1.7ex] \sim
\end{array}}

\newcommand{\dd}{\rmd}
\newcommand{\DD}{D}
\newcommand{\im}{\rmi}
\newcommand{\ee}{\rme}
\newcommand{\perpperp}{\perp\!\!\perp}
\newcommand{\nn}{\nonumber\\}

\newcommand{\Msun}{{\rm M}_\odot}
\newcommand{\vel}{\textsl{v}}
\newcommand{\inn}{{\rm in}}
\newcommand{\out}{{\rm ou}}
\newcommand{\abh}{a_\bullet}
\newcommand{\Mbh}{M_\bullet}
\newcommand{\Mbhdot}{\dot{M}_\bullet}
\newcommand{\Qbh}{Q_\bullet}

\newcommand{\bg}{\bm{g}}
\newcommand{\bp}{\bm{p}}
\newcommand{\bx}{\bm{x}}
\newcommand{\QCarter}{{\cal Q}}
\newcommand{\rc}{{\scriptstyle R}}
\newcommand{\tc}{{\scriptstyle T}}
\newcommand{\smallrc}{{\scriptscriptstyle R}}
\newcommand{\smalltc}{{\scriptscriptstyle T}}

\newcommand{\plus}{{\scriptscriptstyle +}}
\newcommand{\minus}{{\scriptscriptstyle -}}
\newcommand{\plusminus}{{\scriptscriptstyle \pm}}
\newcommand{\minusplus}{{\scriptscriptstyle \mp}}
\newcommand{\zero}{{\scriptstyle 0}}
\newcommand{\one}{{\scriptstyle 1}}
\newcommand{\two}{{\scriptstyle 2}}
\newcommand{\three}{{\scriptstyle 3}}
\newcommand{\comma}{{\scriptscriptstyle ,}}
\newcommand{\smallzero}{{\scriptscriptstyle 0}}
\newcommand{\smallone}{{\scriptscriptstyle 1}}
\newcommand{\Cz}{{\tilde C}}
\newcommand{\nY}[1]{{}_{#1\,} \! Y}

\hyphenpenalty=3000

\newcommand{\slowkerrvarsfig}{
    \begin{figure}[tbhp!]
    \begin{center}
    \leavevmode
    \includegraphics[scale=.75]{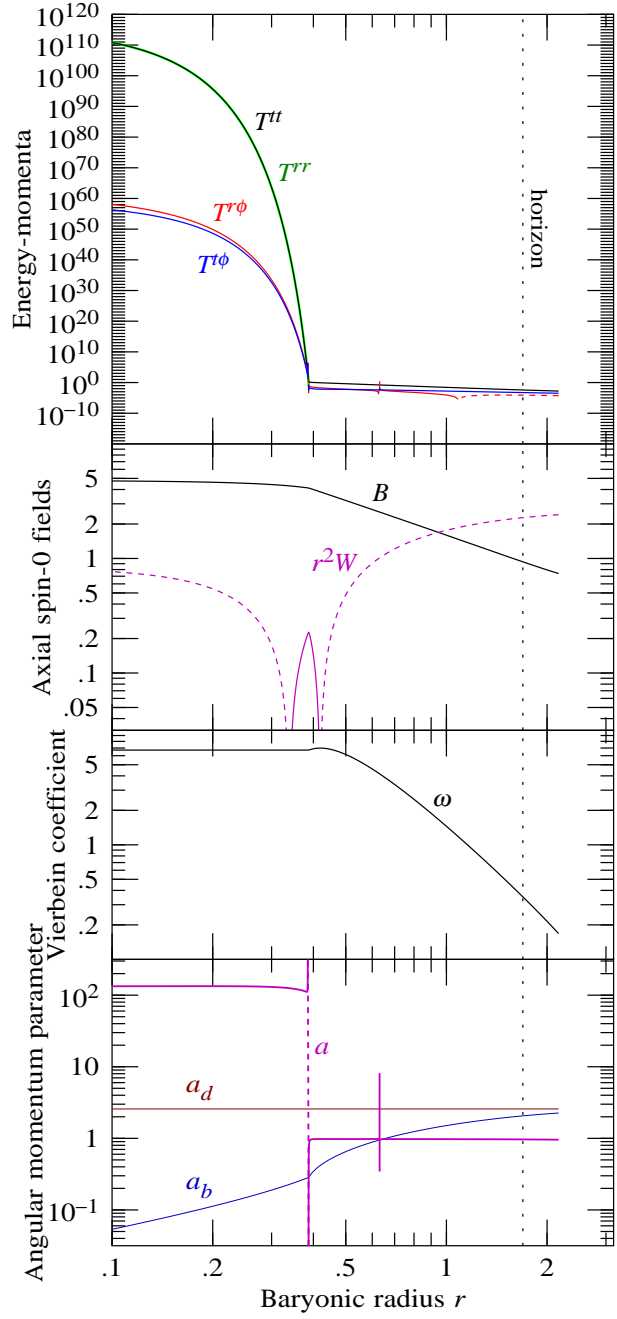}
    \caption[1]{
    \label{slowkerrvars}
Self-similar model with
accretion rate $\Mbhdot = 0.01$,
charge-to-mass $\Qbh/\Mbh = 0.8$,
and
dark-matter-to-baryon density ratio
$\rho_d / \rho_b = 0.1$,
at the sonic point.
Lines are dashed where a quantity is negative.
All quantities are in geometric units,
$c = G = \Mbh = 1$.
Perturbed quantities
(everything except $T^{tt}$ and $T^{rr}$)
are in units where $\abh = 1$,
and should be multiplied by the actual $\abh$,
a small number.
(Top)
Components of the tetrad-frame energy-momentum tensor $T^{mn}$
in a frame that is center-of-mass
in the radial direction
and principal in the angular direction.
(Upper middle)
Axial spin-$0$ Weyl and electromagnetic fields $W$ and $B$,
equations~(\protect\ref{W}) and (\protect\ref{Br}).
(Lower middle)
Vierbein coefficient
$\omega \Mbhdot / \Mbh$
[the scaling adjusts
from conformal time to Kerr-Newman time
at the sonic point, eq.~(\protect\ref{tctkn})].
(Bottom)
Angular momentum parameters
$a_b$ and $a_d$ of the baryons and dark matter,
equation~(\protect\ref{a}),
and $a$ of the black hole in the principal frame,
equation~(\protect\ref{aprincipal}).
    }
    \end{center}
    \end{figure}
}

\newcommand{\Brstarfig}{
    \begin{figure}[tbp!]
    \begin{center}
    \leavevmode
    \includegraphics[scale=.75]{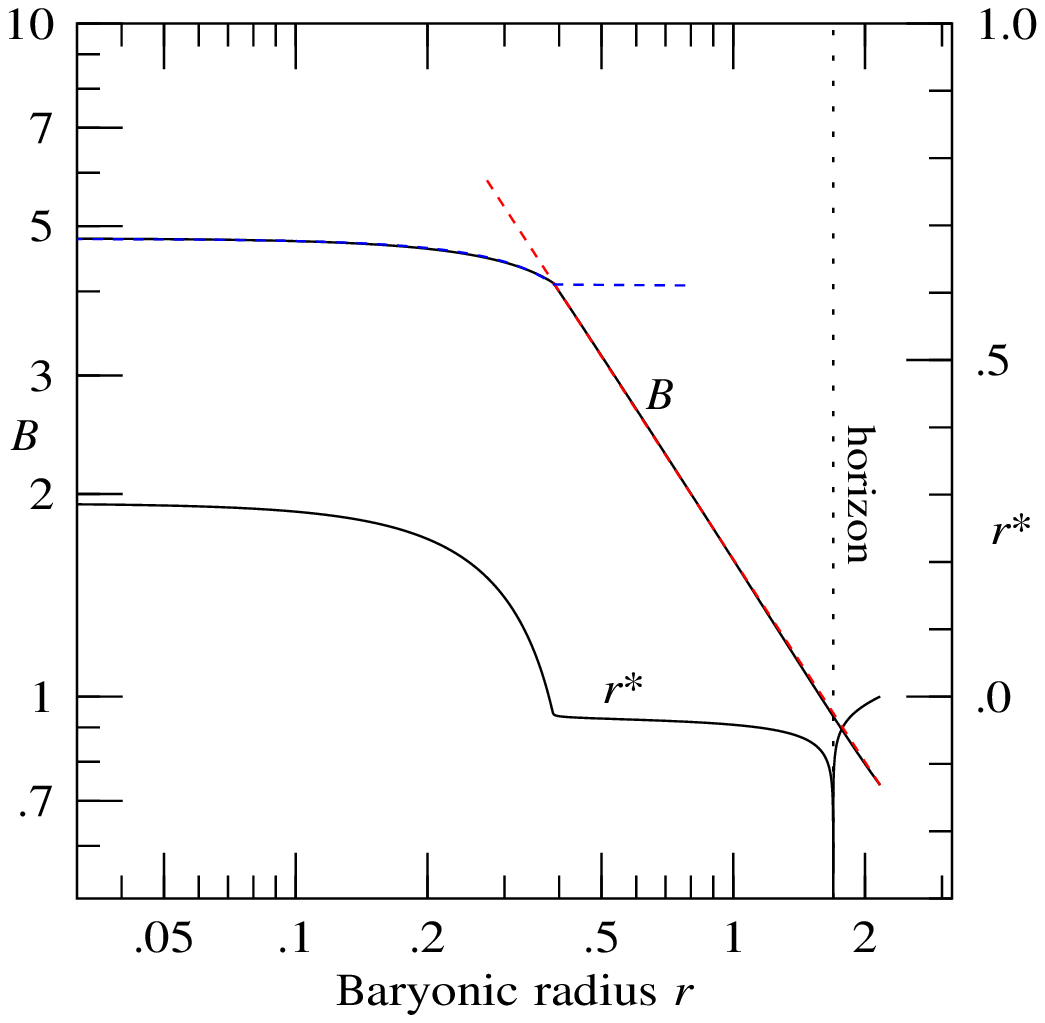}
    \caption[1]{
    \label{Brstar}
Radial magnetic field $B$
(left scale),
and Regge-Wheeler coordinate $r^\ast$
(right scale)
for the model shown in Figure~\protect\ref{slowkerrvars}.
The dashed lines superposed on the magnetic field $B$
show approximations valid respectively in the Kerr-Newman (red)
and inflationary (blue)
regimes,
equations~(\protect\ref{knB}) and (\protect\ref{Binflationapproximation}).
    }
    \end{center}
    \end{figure}
}

\begin{abstract}
The internal structure of a slowly rotating, charged black hole
that is undergoing mass inflation at its inner horizon is derived.
The equations governing the angular behavior decouple from the radial behavior,
so all conclusions regarding inflation in a spherical charged black hole
carry through unchanged for a slowly-rotating black hole.
Quantities inflate only in the radial direction, not in the angular direction.
Exact self-similar solutions are obtained.
For
sufficiently small accretion rates,
the instantaneous angular motion of the accretion flow has
negligible effect on the angular spacetime structure of the black hole,
even if the instantaneous angular momentum of the accretion flow is large
and arbitrarily oriented.
\end{abstract}

\pacs{04.20.-q}	

\date{\today}

\maketitle

\section{Introduction}

The internal structure of a realistic (astronomical) rotating black hole
has remained an unsolved problem for decades
(e.g.\ \S20.5 of \cite{SKMHH03}, \S5.3 of \cite{Joshi07}).
The structure of a rotating black hole in the process of collapsing
is undoubtedly thoroughly complicated
\cite{Berger02}.
However,
thanks to the no-hair theorem
\cite{Israel67,Carter71,Wald71},
it is expected that a newly collapsed rotating black hole will quickly relax
to the Kerr geometry.
Aside from the initial collapse event,
and rare occasions of high accretion such as a black hole merger,
a real astronomical black hole
spends the great majority of its life accurately
described by the Kerr geometry down to near its inner horizon.

The reason that the Kerr geometry fails inside a black hole is that
at (or rather just above) the inner horizon,
the Kerr geometry is subject to the relativistic counter-streaming instability
discovered by
\cite{PI90},
and called by them mass inflation.
The inflationary instability is the nonlinear realization
of the infinite blueshift at the inner horizon first pointed out by
\cite{Penrose68}.
Mass inflation inside spherical charged black holes
has been investigated extensively
(see e.g.\ the lists of references in
\cite{HKN05,HP05b}),
but there have been few studies of inflation inside
rotating black holes
\cite{Ori92,Ori99,BC95,BDM98}
(see also the review \cite{Brady99}).

The present paper addresses, and solves,
the problem of the internal structure of a slowly rotating,
charged black hole
that is undergoing inflation at its inner horizon.
Inflation requires the simultaneous presence of ingoing and outgoing
streams near the inner horizon;
but even the tiniest incident streams suffice to induce inflation.
The paper follows earlier papers in this series
\cite{HP05a,HP05b,HA08}
in assuming that ingoing and outgoing streams are being
generated continuously and steadily by accretion.
This assumption differs from most of the literature on inflation,
which has focused on the situation where a black hole collapses
and then remains isolated
\cite{Dafermos04a},
in which case
the outgoing stream is dominated by
a decaying Price tail
\cite{Price72,DR05}
of gravitational radiation.
Real astronomical black holes are however never isolated,
and it seems likely that any decaying Price tail
produced by the initial collapse event
will soon be overwhelmed by streams generated by accretion.
Matter that falls from outside the horizon
is necessarily initially ingoing,
but it may become outgoing inside the horizon
if it has enough charge (if the black hole is charged)
or enough angular momentum (if the black hole is rotating).

The approach followed in the present paper
is to introduce a small rotational perturbation
to a spherically symmetric black hole that is undergoing inflation.
It is necessary to assume that the black hole is charged
so that the unperturbed geometry has an inner horizon
where inflation is taking place.
In this sense the solution is not realistic,
since a real black hole probably has tiny charge.
However, one may hope that the solution
for the internal structure of a slowly rotating, charged black hole
may shed light on the real problem of a fully rotating,
uncharged black hole.

As shown by \cite{HA08},
the radial length scale over which the interior mass
and other gauge-invariant quantities
(such as the Weyl curvature,
and the center-of-mass density and pressure)
exponentiate during inflation
is proportional to the accretion rate of the incident
ingoing and outgoing streams.
Thus, counter-intuitively,
the smaller the accretion rate,
the more rapidly inflation exponentiates
as a function of radius.
In the limit of a tiny accretion rate,
the structure resembles a step function at the inner horizon,
above which the Kerr-Newman geometry applies to an excellent approximation,
and at which the interior mass
and other quantities inflate to absurdly huge values
while the radius scarcely changes.

Astronomically realistic black holes have small accretion rates
over the great majority of their lifetimes,
the accretion timescale being much longer than the black hole crossing time.
For small accretion rates
it might seem justifiable to assume the stationary approximation
$\partial / \partial t = 0$
(which
\cite{Burko97a,Burko98,Burko99}
terms the homogeneous approximation,
since the Killing vector associated with time translation symmetry
is spacelike inside the horizon).
It is true that the spacetime is almost stationary,
temporal partial derivatives
being much smaller than radial partial derivatives,
$\partial / \partial t \ll \partial / \partial r$,
but it is strictly stationary only in the unrealistic case
of symmetrically equal ingoing and outgoing streams
\cite{HA08}.
The reason that stationarity breaks down
is that quantities such as the interior mass $M$
inflate so rapidly
that $\partial M / \partial t$ is never small
in the inflationary zone
even in the limit of infinitesimal accretion rates.

Although the slowly-rotating black hole spacetimes
considered in this paper are not time-translation invariant,
the equations do admit solutions with
conformal time-translation symmetry,
or self-similarity.
Self-similar solutions are considered herein.

\section{Slowly rotating equations}
\label{slowlyrotating}

\subsection{Slowly rotating perturbation}
\label{slowlyrotatingperturbation}

This paper follows~\cite{HP05a,HP05b,HA08}
in adopting the following line-element
for a general spherically symmetric background spacetime,
in polar coordinates
$x^\mu \equiv \{ t , r , \theta , \phi \}$:
\begin{equation}
\label{lineelement}
  \dd s^2
  =
  - \,
  {\dd t^2 \over \alpha^2}
  +
  {1 \over \beta_r^2}
  \left(
  \dd r
  -
  \beta_t
  {\dd t \over \alpha}
  \right)^2
  +
  r^2
  ( \dd \theta^2 + \sin^2\!\theta \, \dd \phi^2 )
  \ .
\end{equation}
Through the identity
\begin{equation}
  \dd s^2
  =
  g_{\mu\nu}
  \, \dd x^\mu \dd x^\nu
  =
  \eta_{mn}
  e^m{}_\mu e^n{}_\nu
  \, \dd x^\mu \dd x^\nu
  \ ,
\end{equation}
the line-element~(\ref{lineelement}) encodes not only a metric
$g_{\mu\nu}$,
but a complete inverse vierbein $e^m{}_\mu$,
corresponding vierbein $e_m{}^\mu$,
and orthonormal tetrad
$\{ \bgamma_t , \bgamma_r , \bgamma_\theta , \bgamma_\phi \}$,
satisfying
$\bgamma_m \cdot \bgamma_n \equiv \eta_{mn}$
with
$\eta_{mn}$ the Minkowski metric.
The coefficients $\beta_m$
in the line-element~(\ref{lineelement})
constitute the components of
a covariant tetrad-frame 4-vector, the radial 4-gradient
\begin{equation}
\label{betam}
  \beta_m
  \equiv
  \partial_m r
  \ ,
\end{equation}
where
$\partial_m \equiv \bgamma_m \cdot \bpartial \equiv e_m{}^\mu \partial / \partial x^\mu$
denotes the directed derivative along the
$\bgamma_m$
tetrad axis
(not to be confused with the coordinate derivative
$\partial / \partial x^\mu$).
Physically,
the directed derivatives
$\partial_t$
and
$\partial_r$
are the proper time and radial derivatives
measured by a person at rest in the tetrad frame.
The scalar product of $\beta_m$ with itself defines the interior,
or Misner-Sharp \cite{MS64},
mass $M$,
a gauge-invariant scalar
\begin{equation}
  {2 M \over r} - 1
  \equiv
  - \beta_m \beta^m
  =
  \beta_t^2 - \beta_r^2
  \ .
\end{equation}
Details of the vierbein,
tetrad-frame connections,
and tetrad-frame Riemann, Einstein, and Weyl tensors
that follow from the unperturbed line-element~(\ref{lineelement})
are summarized in Appendix~A of \cite{HA08}.

Perturbations to the spherically symmetric spacetime
can be characterized by the
dimensionless covariant tetrad-frame vierbein perturbation $\varphi_{mn}$
defined by
\begin{equation}
  e_m{}^\mu
  =
  ( \delta_m^n - \varphi_m{}^n )
  \overset{\smallzero}{e}_n{}^\mu
  \ ,
\end{equation}
where the overscript $^{\smallzero}$
signifies an unperturbed quantity.
Because the perturbation $\varphi_{mn}$
is of linear order,
it is legitimate to raise and lower its indices
with the unperturbed tetrad-frame metric
(the Minkowski metric),
and to transform between coordinate and tetrad frames
with the unperturbed vierbein
$\overset{\smallzero}{e}_m{}^\mu$
and its inverse
$\overset{\smallzero}{e}{}^m{}_\mu$.
Thus the vierbein perturbation $\varphi_{mn}$
can be regarded as a tensor field defined on the unperturbed
spherically symmetric background.

Spherically symmetric spacetimes
are invariant under the 3-dimensional group $O(3)$
of orthogonal transformations.
It is natural to expand perturbations in eigenmodes of that group.
A perturbation eigenmode
is characterized not only by its harmonic numbers $lm$,
but also by its spin $s$,
which expresses how the perturbation transforms
under rotations about the radial direction.
Perturbations $\varphi_{mn}$
admit scalar (spin-$0$), vector (spin-$\pm 1$),
and tensor (spin-$\pm 2$) modes.
The spin of a perturbation
$\varphi_{mn}$
is most easily identified not in an orthonormal tetrad,
as here,
but rather in a spinor tetrad,
where the angular tetrad axes
$\bgamma_\theta$ and $\bgamma_\phi$ are replaced by spinor axes
\begin{equation}
  \bgamma_\plus
  \equiv
  {1 \over \sqrt{2}}
  ( \bgamma_\theta + \im \bgamma_\phi )
  \ , \quad
  \bgamma_\minus
  \equiv
  {1 \over \sqrt{2}}
  ( \bgamma_\theta - \im \bgamma_\phi )
  \ .
\end{equation}
The general rule is that the spin of a tensor
equals the sum of the $+$'s and $-$'s of its covariant indices
in the spinor tetrad.

A slowly rotating spacetime is
obtained from a spherically symmetric spacetime by
a perturbation that is
(a) cylindrically symmetric,
(b) axial,
meaning that the perturbation changes sign under a flip
$\bgamma_\phi \rightarrow - \bgamma_\phi$
of the azimuthal tetrad axis,
and
(c) dipole,
the lowest order non-spherical harmonic.
The slowly rotating conditions
imply that the vierbein perturbation
$\varphi_{mn}$
is a spin-$1$ dipole harmonic,
\begin{equation}
  \varphi_{mn}
  \propto
  \nY{\one}_{\one\zero}(\theta,\phi)
  =
  - {\sin\theta \over \sqrt{8\pi}}
  \ .
\end{equation}
Of the 16 components of the $4 \times 4$ tensor of perturbations
$\varphi_{mn}$,
there are 4 that satisfy the conditions of being spin-$1$ and axial, namely
\begin{equation}
\label{varphi}
  \varphi_{t\phi}
  \ , \quad
  \varphi_{r\phi}
  \ , \quad
  \varphi_{\phi t}
  \ , \quad
  \varphi_{\phi r}
  \ .
\end{equation}
The spin-$1$ axial character of these perturbations is
manifest in a spinor tetrad,
where the possible spin-$1$ perturbations are the 4 complex quantities
$\varphi_{t\plus}$,
$\varphi_{r\plus}$,
$\varphi_{\plus t}$,
and
$\varphi_{\plus r}$,
and the axial condition implies that each of these
4 perturbations is purely imaginary.

Of the 4 possible degrees of freedom
represented by the perturbations~(\ref{varphi}),
only one degree of freedom is coordinate and tetrad gauge-invariant,
and therefore represents a real physical perturbation.
The remaining 3 degrees of freedom represent gauge freedoms,
comprising
an infinitesimal coordinate transformation
of the azimuthal coordinate $\phi$,
and infinitesimal tetrad (Lorentz) transformations
in the $t$-$\phi$ and $r$-$\phi$ planes.

Under an infinitesimal coordinate gauge transformation
of the azimuthal coordinate $\phi$,
\begin{equation}
  \phi \rightarrow \phi + \epsilon
  \ ,
\end{equation}
the vierbein perturbations transform as
\numparts
\begin{eqnarray}
  \varphi_{t\phi}
  &\rightarrow&
  \varphi_{t\phi}
  -
  r \sin\theta
  \left(
  \alpha
  {\partial \epsilon \over \partial t}
  +
  \beta_t
  {\partial \epsilon \over \partial r}
  \right)
  \ ,
\\
  \varphi_{r\phi}
  &\rightarrow&
  \varphi_{r\phi}
  -
  r \sin\theta \,
  \beta_r {\partial \epsilon \over \partial r}
  \ ,
\end{eqnarray}
\endnumparts
with
$\varphi_{\phi t}$ and $\varphi_{\phi r}$
being coordinate gauge-invariant.
If the spacetime were either time translation-invariant
or conformally time translation-invariant,
so that
$\partial \epsilon / \partial t = 0$,
then the combination
\begin{equation}
\label{coordinateinvariantperturbation}
  \beta_r
  \varphi_{t\phi}
  -
  \beta_t
  \varphi_{r\phi}
\end{equation}
would be coordinate gauge-invariant.
In general the spacetime is not time translation-invariant,
conformal or otherwise,
but as commented in the Introduction
it is legitimate to think of the spacetime as being almost stationary,
and therefore the combination~(\ref{coordinateinvariantperturbation})
is almost coordinate gauge-invariant.

The antisymmetric combinations
$\varphi_{t\phi} - \varphi_{\phi t}$
and
$\varphi_{r\phi} - \varphi_{\phi r}$
are not tetrad gauge-invariant,
the former transforming under an infinitesimal Lorentz boost
in the $t$-$\phi$ plane,
the latter transforming under an infinitesimal spatial rotation
in the $r$-$\phi$ plane.
The symmetric combinations
$\varphi_{t\phi} + \varphi_{\phi t}$
and
$\varphi_{r\phi} + \varphi_{\phi r}$
are tetrad gauge-invariant.
If the spacetime
were time translation-invariant
or conformally time translation-invariant
then the combination
\begin{equation}
\label{gaugeinvariantperturbation}
  \beta_r ( \varphi_{t\phi} + \varphi_{\phi t} )
  -
  \beta_t
  ( \varphi_{r\phi} + \varphi_{\phi r} )
\end{equation}
would be
gauge-invariant with respect to both infinitesimal coordinate and tetrad
gauge transformations.
Again,
in general the spacetime is not time translation-invariant, but
it is legitimate to think of the combination~(\ref{gaugeinvariantperturbation})
as being almost coordinate and tetrad gauge-invariant.
In terms of the almost-gauge-invariant perturbation $\omega$
defined below, the combination~(\ref{gaugeinvariantperturbation}) equals
$- r \alpha \beta_r \omega \sin\theta$.
In the general case,
there is no linear combination of vierbein perturbations
$\varphi_{mn}$
that is fully gauge-invariant,
but the axial spin-$0$ component of the Weyl tensor
provides a differential combination $W$ of vierbein perturbations,
equation~(\ref{W}),
that is fully gauge-invariant.

In view of the behavior under gauge transformations just discussed,
it is convenient to introduce
4 quantities
$\omega$, $\psi$,
$\zeta^t$, and $\zeta^r$
representing the different types of perturbation as follows:
\numparts
\label{perturbationvariables}
\begin{eqnarray}
\label{omega}
  \omega
  &\ 
  \mbox{almost gauge-invariant}
  \ ,
\\
\label{psi}
  \psi
  &\ 
  \mbox{coordinate transformation of $\phi$}
  \ ,
\\
\label{zetat}
  \zeta^t
  &\ 
  \mbox{Lorentz boost in the $t$-$\phi$ plane}
  \ ,
\\
\label{zetab}
  \zeta^r
  &\ 
  \mbox{spatial rotation in the $r$-$\phi$ plane}
  \ ,
\end{eqnarray}
\endnumparts
in terms of which
the vierbein perturbation $\varphi_{mn}$ is
\begin{equation}
\label{perturbation}
  \varphi_{mn}
  =
  \sin\theta
  \left(
  \begin{array}{cccc}
    0 & 0 & 0 &
    \displaystyle
    - r \alpha \omega
    - \beta_t \psi
    + \zeta^t
    \\
    0 & 0 & 0 &
    \displaystyle
    - \beta_r \psi
    - \zeta^r
    \\
    0 & 0 & 0 & 0 \\
    \displaystyle
    - \zeta^t
    &
    \displaystyle
    \zeta^r
    & 0 & 0
  \end{array}
  \right)
  \ .
\end{equation}
The perturbations
$\zeta^m$
are tetrad-frame components of
the Killing vector associated with cylindrical symmetry,
equation~(\ref{zetam}).
The resulting perturbed vierbein is
\begin{equation}
\label{perturbedvierbein}
  e_m{}^\mu
  =
  \left(
  \begin{array}{cccc}
    \alpha & \beta_t & 0 &
    \displaystyle
    \alpha \Omega + \beta_t \Psi \\
    0 & \beta_r & 0 & \beta_r \Psi \\
    0 & 0 &
    \displaystyle
    {1 \over r} & 0 \\
    \displaystyle
    - \zeta^t \alpha \sin \theta
    &
    \displaystyle
    \beta_\phi
    \ & 0 &
    \displaystyle
    {1 \over r \sin\theta}
  \end{array}
  \right)
  \ ,
\end{equation}
and the perturbed inverse vierbein is
\begin{equation}
\label{perturbedinversevierbein}
  e^m{}_\mu
  =
  \left(
  \begin{array}{cccc}
    \displaystyle
    {1 \over \alpha}
    & 0 & 0 &
    \displaystyle
    r \zeta^t \sin^2\!\theta
    \\
    \displaystyle
    - {\beta_t \over \alpha \beta_r}
    &
    \displaystyle
    {1 \over \beta_r} & 0 &
    \displaystyle
    r \zeta^r \sin^2\!\theta
    \\
    0 & 0 &
    \displaystyle
    r & 0 \\
    \displaystyle
    - r \Omega \sin\theta &
    \displaystyle
    - \Psi \sin\theta & 0 &
    \displaystyle
    r \sin\theta
  \end{array}
  \right)
  \ ,
\end{equation}
where
\numparts
\begin{eqnarray}
\label{Omega}
  \Omega
  &\equiv&
  \omega
  - {\zeta^t \beta_r + \zeta^r \beta_t \over r \alpha \beta_r}
  \ ,
\\
\label{Psi}
  \Psi
  &\equiv&
  \psi
  +
  {\zeta^r \over \beta_r}
  \ .
\end{eqnarray}
\endnumparts

The final column of the inverse vierbein~(\ref{perturbedinversevierbein})
constitutes the
spacelike contravariant tetrad-frame Killing vector
$r \sin^2\!\theta \, \zeta^m$
associated with cylindrical symmetry
\begin{equation}
\label{zetam}
  {\partial \over \partial \phi}
  =
  e^m{}_\phi \, \partial_m
  \equiv
  r \sin^2\!\theta \, \zeta^m \partial_m
  \ .
\end{equation}

The second column of the vierbein~(\ref{perturbedvierbein})
constitutes the
covariant tetrad-frame 4-vector $\beta_m$,
the radial 4-gradient defined by equation~(\ref{betam}).
The azimuthal component $\beta_\phi$ of the 4-vector follows from
$\partial r / \partial \phi = 0$,
or equivalently
$\zeta^m \beta_m = 0$,
which implies
\begin{equation}
\label{betaphi}
  \beta_\phi
  =
  - \left( \zeta^t \beta_t + \zeta^r \beta_r \right) \sin\theta
  \ .
\end{equation}

The inverse vierbein~(\ref{perturbedinversevierbein}) is equivalent to
the line-element
\begin{eqnarray}
\label{perturbedlineelement}
  \dd s^2
  =&
  r^2 \biggl[
  - \,
  \left(
  {\dd t \over r \alpha}
  + \zeta^t
  \sin^2\!\theta
  \, \dd \phi
  \right)^2
  +
  \left(
  {\dd \ln r \over \beta_r}
  -
  {\beta_t \over \beta_r} {\dd t \over r \alpha}
  + \zeta^r \sin^2\!\theta \, \dd \phi
  \right)^2
\nn
  &
  + \,
  \dd \theta^2
  +
  \sin^2\!\theta
  \left( \dd \phi - \Psi \, \dd \ln r - \Omega \, \dd t \right)^2
  \biggr]
  \ .
  \qquad
\end{eqnarray}

\subsection{Einstein and Weyl tensors}
\label{EinsteinWeyl}

From the line-element~(\ref{perturbedlineelement})
can be derived the tetrad-frame connections,
and tetrad-frame Riemann, Ricci, Einstein, and Weyl tensors
in the usual way.
In the present case of a spherical spacetime
subjected to an axial vector perturbation,
all quantities can be classified as
either scalar (spin-$0$) or vector (spin-$1$).

The only spin-$0$ component of the Riemann tensor
that differs from its value in the unperturbed spherical background
is the axial part of the spin-$0$ component of the Weyl tensor
$C_{klmn}$.
The component is proportional to the spin-$0$ dipole harmonic
$\nY{0}_{10} = \cos\theta / \sqrt{4\pi}$,
and is conveniently written
\begin{equation}
\label{axialWeylscalar}
  {W \cos\theta \over r^2}
  \equiv
  -
  \frac{1}{2}
  C_{tr\theta\phi}
  =
  -
  C_{t\theta r\phi}
  =
  C_{t\phi r\theta}
  \ ,
\end{equation}
where
$W(t,r)$ is a function only of $t$ and $r$,
not of angular variables.
In terms of the vierbein perturbations,
$W$ is explicitly
\begin{equation}
\label{W}
  W
  =
  {r \alpha \beta_r \over 2}
  \left(
  {\partial \omega \over \partial \ln r}
  -
  {\partial \psi \over \partial t}
  \right)
  \ .
\end{equation}
The axial spin-$0$ component of the Weyl tensor,
equation~(\ref{axialWeylscalar}),
hence $W$,
is fully gauge-invariant
with respect to all relevant gauge transformations,
which is to say not only
infinitesimal coordinate and tetrad transformations,
but also arbitrary radial Lorentz boosts.

There are 4 non-vanishing axial spin-$1$ components of the
tetrad-frame Riemann tensor
$R_{klmn}$,
yielding the following 4 non-vanishing spin-$1$ components
of the tetrad-frame Einstein tensor $G_{mn}$
and Weyl tensor
$C_{klmn}$:
\numparts
\begin{eqnarray}
  G_{t\phi}
  =
  -
  \left(
  R_{t\theta\theta\phi}
  +
  R_{trr\phi}
  \right)
  \ ,
\\
  G_{r\phi}
  =
  -
  \left(
  R_{r\theta\theta\phi}
  +
  R_{trt\phi}
  \right)
  \ ,
\\
  C_{t\theta\theta\phi}
  =
  -
  C_{trr\phi}
  =
  \frac{1}{2}
  \left(
  R_{t\theta\theta\phi}
  -
  R_{trr\phi}
  \right)
  \ ,
\\
  C_{r\theta\theta\phi}
  =
  -
  C_{trt\phi}
  =
  \frac{1}{2}
  \left(
  R_{r\theta\theta\phi}
  -
  R_{trt\phi}
  \right)
  \ .
\end{eqnarray}
\endnumparts
Being tetrad-frame quantities,
all these components are automatically coordinate gauge-invariant.
However, none of them are tetrad gauge-invariant:
all spin-$1$ components of the Riemann tensor transform
under infinitesimal spin-$1$ tetrad (Lorentz) transformations.

The non-vanishing spin-$1$ components of the
tetrad-frame Einstein tensor $G_{mn}$ are
\numparts
\label{Gphi}
\begin{eqnarray}
\label{Gtphi}
  G_{t\phi}
  =
  \sin\theta
  \left[
  {1 \over r^3}
  \partial_r ( r^2 W )
  - \zeta^t
  (
  G_{tt}
  +
  G_{\phi\phi}
  )
  - \zeta^r
  G_{tr}
  \right]
  \ ,
\\
\label{Grphi}
  G_{r\phi}
  =
  \sin\theta
  \left[
  {1 \over r^3}
  \partial_t ( r^2 W )
  - \zeta^r
  (
  G_{rr}
  -
  G_{\phi\phi}
  )
  - \zeta^t
  G_{tr}
  \right]
  \ .
\end{eqnarray}
\endnumparts
Note that the derivatives $\partial_m$ in
equations~(24)
are directed (tetrad-frame) derivatives,
not partial (coordinate-frame) derivatives.

The non-vanishing spin-$1$ components of the
tetrad-frame Weyl tensor
$C_{klmn}$ are
\numparts
\label{Cphi}
\begin{eqnarray}
  C_{t\theta\theta\phi}
  =
  \sin\theta
  \left(
  {1 \over 2 r}
  \partial_r W
  -
  3
  \zeta^t
  C
  \right)
  \ ,
\\
  C_{r\theta\theta\phi}
  =
  \sin\theta
  \left(
  {1 \over 2 r}
  \partial_t W
  +
  3
  \zeta^r
  C
  \right)
  \ ,
  \quad
\end{eqnarray}
\endnumparts
where
$C$ is the Weyl scalar,
the polar part of the spin-$0$ component of the Weyl tensor
\begin{eqnarray}
\label{C}
  C
  &\equiv
  \frac{1}{2}
  C_{trtr}
  =
  -
  C_{t\theta t\theta}
  =
  -
  C_{t\phi t\phi}
  =
  -
  C_{r\theta r\theta}
  =
  -
  C_{r\phi r\phi}
\nn
  &=
  -
  \frac{1}{2}
  C_{\theta\phi\theta\phi}
  =
  {1 \over 6} ( G_{tt} - G_{rr} + G_{\phi\phi} ) - {M \over r^3}
  \ .
\end{eqnarray}
The Weyl scalar
$C$
is invariant with respect to arbitrary radial Lorentz boosts.

\subsection{Electromagnetic field}

The black hole is necessarily charged,
in order that the unperturbed black hole have an inner horizon
where inflation can take place.

The unperturbed spherical background admits just one component
of the electromagnetic field
$F_{mn}$,
the polar spin-$0$ component,
the radial electric field $E_r \equiv F_{tr}$.
This is a monopole field,
and it can be written
\begin{equation}
\label{Er}
  E_r
  \equiv
  {Q \over r^2}
  \ ,
\end{equation}
which defines the charge $Q(t,r)$ interior to radius $r$.

The rotational perturbation introduces
scalar (spin-$0$) and vector (spin-$1$) axial dipole components
to the electromagnetic field.
The axial spin-$0$ part is the radial magnetic field
$B_r \equiv F_{\phi\theta}$.
The component is proportional to the spin-$0$ dipole harmonic
$\nY{0}_{10} = \cos\theta / \sqrt{4\pi}$,
and can be written
\begin{equation}
\label{Br}
  B_r
  \equiv
  {B \cos\theta \over r^2}
  \ ,
\end{equation}
which defines $B(t,r)$.
Both $Q$ and $B$,
equations~(\ref{Er}) and (\ref{Br}),
are fully gauge-invariant
with respect to infinitesimal coordinate and tetrad gauge transformations,
and to arbitrary radial Lorentz boosts.

The non-vanishing axial spin-$1$ components of the electromagnetic field are
$E_\phi \equiv F_{t\phi}$
and
$B_\theta \equiv F_{r\phi}$.
The source-free Maxwell's equations
$\varepsilon^{klmn} D_l F_{mn} = 0$
imply that
the axial spin-$1$ components
of the electromagnetic field must be related to $Q$ and $B$ by
\numparts
\label{EphiBtheta}
\begin{eqnarray}
  E_\phi
  =
  -
  \sin\theta
  \left(
  \frac{1}{2 r}
  \partial_t B
  + \zeta^r
  {Q \over r^2}
  \right)
  \ ,
\\
  B_\theta
  =
  -
  \sin\theta
  \left(
  \frac{1}{2 r}
  \partial_r B
  - \zeta^t
  {Q \over r^2}
  \right)
  \ .
\end{eqnarray}
\endnumparts

The source Maxwell's equations
$D^m F_{mn} = 4 \pi j_n$
relating the electromagnetic field $F_{mn}$
to the electric current $j_n$
are
\numparts
\label{j}
\begin{eqnarray}
\label{jt}
  \partial_r Q
  =
  - 4 \pi r^2 j_t
  \ ,
\\
\label{jr}
  \partial_t Q
  =
  - 4 \pi r^2 j_r
  \ ,
\\
  \frac{1}{2}
  \left(
  -
  D^t \partial_t - D^r \partial_r
  +
  {2 \over r^2}
  \right)
  B
  +
  {2 Q W \over r^2}
  =
  4 \pi
  r \zeta^m j_m
  \ ,
  \quad\quad\ 
\label{jphi}
\end{eqnarray}
\endnumparts
where $D_m$ signifies the tetrad-frame covariant derivative.

The electromagnetic energy-momentum tensor is given by
$4 \pi T^{e}_{mn} = F_{mk} F_n{}^k - \frac{1}{4} \eta_{mn} F_{kl} F^{kl}$.
In terms of the electric and magnetic fields~(\ref{Er}) and
(29),
the axial spin-$1$ components of
the electromagnetic energy-momentum tensor
are
\begin{equation}
\label{Tephi}
  4 \pi T^{e}_{t\phi} = - E_r B_\theta
  \ , \quad
  4 \pi T^{e}_{r\phi} = - E_r E_\phi
  \ .
\end{equation}
Explicitly,
from
equations~(29)
inserted into equations~(\ref{Tephi}),
\numparts
\label{TephiB}
\begin{eqnarray}
  8 \pi \, T^{e}_{t\phi}
  =
  \sin\theta
  \left(
  {Q \over r^3}
  \partial_r B
  + \zeta^t
  {2 Q^2 \over r^2}
  \right)
  \ ,
\\
  8 \pi \, T^{e}_{r\phi}
  =
  \sin\theta
  \left(
  {Q \over r^3}
  \partial_t B
  - \zeta^r
  {2 Q^2 \over r^2}
  \right)
  \ .
  \quad
\end{eqnarray}
\endnumparts
The covariant divergence of the electromagnetic energy-momentum is
$D_m T_{e}^{mn} = F^{nm} j_m$,
and the component of this along the azimuthal Killing vector $\zeta^n$ is
\begin{equation}
\label{zetaDTe}
  \zeta_n
  D_m T_{e}^{mn}
  =
  \frac{1}{2 r}
  j^m \partial_m B
  \ .
\end{equation}

\subsection{Perfect fluid streams}

The inflation instability requires the simultaneous presence
of ingoing and outgoing streams near the inner horizon
\cite{PI90}.
I follow \cite{HP05b,HA08}
in considering two separate streams,
charged ``baryons'' with a relativistic equation of state
(a value slightly less than $1/3$
allows for the expected slow increase with temperature
of the effective number of relativistic species),
\begin{equation}
\label{wb}
  w \equiv
  p_b / \rho_b = 0.32
  \quad
  (\mbox{baryons})
  \ ,
\end{equation}
and neutral ``dark matter''
with a pressureless equation of state,
\begin{equation}
\label{wd}
  p_d / \rho_d = 0
  \quad
  (\mbox{dark matter})
  \ .
\end{equation}
Repelled by the charge of the black hole
generated self-consistently by their accretion,
the baryons become outgoing inside the horizon.
The dark matter falls freely into the black hole,
and remains ingoing.
Inflation is produced by relativistic counter-streaming
between the outgoing baryons and the ingoing dark matter.

Each of the two streams constitutes a perfect fluid.
The equations in this subsection are for a general perfect fluid.

The tetrad-frame energy-momentum
$T^{mn}$
of a perfect fluid of proper density and pressure $\rho$ and $p$
moving at tetrad-frame 4-velocity $u^m$ is
\begin{equation}
\label{Tperfect}
  T^{mn}
  =
  ( \rho + p ) \, u^m u^n
  + p \, \eta^{mn}
  \ .
\end{equation}

Since the spacetime is cylindrically symmetric,
any geodesic is characterized by a conserved azimuthal angular momentum,
a gauge-invariant scalar.
The azimuthal angular momentum (per unit mass)
of a stream is given by the covariant azimuthal component
$\upsilon_\phi$
of the coordinate-frame 4-velocity $\upsilon^\mu$
(the coordinate-frame 4-velocity $\upsilon^\mu$
is written with a Greek upsilon
to avoid any possible confusion with
the tetrad-frame 4-velocity $u^m$,
written with a Latin u):
\begin{equation}
\label{Lu}
  \upsilon_{\phi}
  =
  e^m{}_\phi \,
  \eta_{mn} \,
  u^n
  =
  r \sin^2\!\theta \, \zeta_m u^m
  \ .
\end{equation}
It is convenient to rewrite the azimuthal angular momentum $\upsilon_\phi$
of a stream in terms of an angular momentum parameter $a$ defined by
\begin{equation}
\label{a}
  \upsilon_\phi
  \equiv
  a \sin^2\!\theta
  \ .
\end{equation}
If the stream is slowly rotating, and if it has no $\theta$ motion
($u^\theta = 0$),
so that its rest frame belongs to the system of
slowly-rotating tetrads considered in \S\ref{slowlyrotatingperturbation},
then $a = r \zeta_t$ in the tetrad rest-frame of the stream.
However, equations~(\ref{Lu}) and (\ref{a}) remain valid in general,
regardless of the angular motion of the stream.
Equations~(\ref{Lu})
and (\ref{a})
imply that
the azimuthal component $u^\phi$ of the tetrad-frame 4-velocity of a stream
is related to the time and radial components $u^t$ and $u^r$,
and to its angular momentum parameter $a$, by
\begin{equation}
\label{uphi}
  u^\phi
  =
  \sin\theta
  \left(
  {a \over r}
  -
  \zeta_t u^t
  -
  \zeta_r u^r
  \right)
  \ .
\end{equation}
With $u^\phi$ related to $u^t$ and $u^r$
by equation~(\ref{uphi}),
it follows from equation~(\ref{Tperfect})
that the axial spin-$1$ components of the tetrad-frame energy-momentum tensor
of a perfect fluid are
\numparts
\label{Tperfectphi}
\begin{eqnarray}
  T^{t\phi}
  =
  \sin\theta
  \left[
  {a
  ( \rho + p ) u^t \over r}
  - \zeta_t
  \left(
  T^{tt} + T^{\phi\phi}
  \right)
  -
  \zeta_r
  T^{tr}
  \right]
  \ ,
\\
  T^{r\phi}
  =
  \sin\theta
  \left[
  {a
  ( \rho + p ) u^r \over r}
  - \zeta_r
  \left(
  T^{rr} - T^{\phi\phi}
  \right)
  - \zeta_t
  T^{tr}
  \right]
  \ .
\end{eqnarray}
\endnumparts
The component of the covariant divergence of the perfect-fluid energy-momentum
along the azimuthal Killing vector $\zeta^n$ is
\begin{equation}
\fl
  \zeta_n
  D_m T^{mn}
  =
  {1 \over r^3}
  \left\{
  \left( \partial_t + h_r \right)
  \left[
  r^2
  a ( \rho + p ) u^t
  \right]
\label{zetaDT}
  +
  \left( \partial_r + h_t \right)
  \left[
  r^2
  a ( \rho + p ) u^r
  \right]
  \right\}
  \ ,
\end{equation}
where $h_m \equiv \Gamma_{trm}$
are tetrad-frame connection coefficients,
equation~(64) of \cite{HA08}.


\subsection{Einstein equations}
\label{einsteineqs}

The Einstein equations are
\begin{equation}
\label{Einsteinphi}
  G_{mn}
  =
  8\pi T_{mn}
  \ .
\end{equation}
With the Einstein tensor from
equation~(24)
and the electromagnetic and perfect fluid energy-momentum tensors from
equations~(\ref{Tephi}) and
(40),
the axial spin-$1$ Einstein equations~(\ref{Einsteinphi})
reduce to
\numparts
\label{Einsteintogether}
\begin{eqnarray}
\label{Einsteintogethert}
  \partial_r ( r^2 W )
  =
  Q
  \partial_r B
  -
  8\pi
  r^2
  \sum_{\rm streams}
  a
  ( \rho + p ) u^t
  \ ,
\\
\label{Einsteintogetherr}
  \partial_t ( r^2 W )
  =
  Q
  \partial_t B
  +
  8\pi
  r^2
  \sum_{\rm streams}
  a
  ( \rho + p ) u^r
  \ ,
\end{eqnarray}
\endnumparts
where the sum is over the accreting baryonic and dark-matter streams.
These
equations~(43)
are gauge-invariant,
the parts depending on the gauge perturbations
$\zeta^t$ and $\zeta^r$
in
equations~(24),
(\ref{Tephi}), and
(40)
having canceled between the left and right hand sides.

If the underlying sources of energy-momentum
$T_{mn}$
are arranged to satisfy conservation of energy-momentum,
as they should,
then one of the two Einstein
equations~(43)
can be treated as redundant,
since it serves simply to enforce
overall covariant energy-momentum conservation
in the azimuthal direction
\begin{equation}
\label{DGphi}
  \zeta^m
  D^n G_{mn}
  =
  0
  \ .
\end{equation}
That there is effectively only one independent
axial spin-$1$ Einstein equation
is consistent with the fact that there
is only one physical degree of freedom
in the axial spin-$1$ vierbein perturbations.

\subsection{Energy exchange}
\label{energyexchange}

To complete the equations,
it is necessary to specify the exchange of energy between
each pair of species.

I assume
for simplicity
that the dark matter and the baryons
stream freely through each other without interacting.

For the interaction between charged baryons and the electromagnetic field,
I assume Ohm's law
in the rest frame of the baryons,
\begin{equation}
  j_m = \sigma E_m
  \quad
  ( m = r , \phi )
  \ ,
\end{equation}
where $\sigma$
is an electrical conductivity.
The azimuthal current
$r \zeta^m j_m$
that appears on the right hand side of the Maxwell equation~(\ref{jphi})
is then, in the rest frame of the baryons,
\begin{equation}
\label{zetaj}
  r \zeta^m j_m
  =
  -
  \left(
  a j_t
  +
  \frac{\sigma}{2}
  \partial_t B
  \right)
  \ .
\end{equation}

\section{Slowly-rotating Kerr-Newman}
\label{slowlyrotatingrn}

As long as the accretion rate is small,
as is true for astronomically realistic black holes,
the spacetime down to near the inner horizon
is well-approximated by
the slowly-rotating Kerr-Newman
\cite{Kerr63,NCCEPT65}
geometry.
The Kerr-Newman geometry
is characterized by
the mass $\Mbh$, charge $\Qbh$,
and angular momentum parameter $\abh$ of the black hole,
all constants.

The monopole part of
the slowly-rotating Kerr-Newman solution
coincides with Reissner-Nordstr\"om.
The interior mass $M(r)$ and charge $Q(r)$ are
\begin{equation}
  M = \Mbh - {\Qbh^2 \over 2 r}
  \ , \quad
  Q = \Qbh
  \ .
\end{equation}
If the time coordinate $t$
is chosen to reflect the stationarity of the spacetime,
$\partial / \partial t = 0$,
and scaled globally
to synchronize with proper time at rest at infinity,
then the vierbein coefficient $\alpha$
is related to the vierbein coefficients $\beta_m$ by
\begin{equation}
  \alpha
  =
  {1 \over \beta_r}
  \ .
\end{equation}
The radial 4-gradient $\beta_m$, equation~(\ref{betam}),
is a tetrad-frame 4-vector,
and different choices of $\beta_t$ or $\beta_r$
correspond to different gauge choices of radial Lorentz boost
of the tetrad frame.
For example,
$\beta_t = 0$ recovers the standard diagonal
Reissner-Nordstr\"om line-element,
while
$\beta_r = 1$ yields the Gullstrand-Painlev\'e
\cite{HL08}
line-element.

The dipole part of the slowly-rotating Kerr-Newman geometry
is specified by
the gauge-invariant vierbein perturbation $\omega(r)$,
and by the gauge-invariant quantities $W(r)$ and $B(r)$
describing the axial spin-$0$ Weyl and electromagnetic scalars,
equations~(\ref{W}) and (\ref{Br}):
\numparts
\label{kn}
\begin{eqnarray}
\label{knomega}
  \omega
  =
  {2 \abh M \over r^3}
  \ ,
\\
\label{knW}
  W
  =
  {\abh \over r^2}
  \left( - \, 3 \Mbh +  {2 \Qbh^2 \over r} \right)
  \ ,
\\
\label{knB}
  B
  =
  {2 \abh \Qbh \over r}
  \ .
\end{eqnarray}
\endnumparts
The Kerr-Newman geometry is most commonly expressed
in the Boyer-Linquist
\cite{BL67}
tetrad,
where the spin-$1$ components of the
Einstein, Weyl, and electromagnetic tensors all vanish identically.
The Boyer-Linquist tetrad is attained with the gauge choices
\begin{equation}
  r \zeta^t
  =
  - \abh \beta_r
  \ , \quad
  r \zeta^r
  =
  \abh \beta_t
  \ .
\end{equation}
The coordinate gauge perturbation $\psi$, or equivalently $\Psi$,
equation~(\ref{Psi}),
plays no role,
but it simplifies the line-element~(\ref{perturbedlineelement}) to set
\begin{equation}
  \Psi = 0
  \ .
\end{equation}

\section{Self-similar solutions}
\label{similarity}

The slowly-rotating black hole equations
admit self-similar solutions.
Though idealized as a model of reality,
the similarity solutions are exact solutions
that provide insight into the general case.
The similarity solutions in this section
are extensions of those discussed by
\cite{HP05a,HP05b,HA08}.

As has been seen in \S\ref{slowlyrotating},
the equations governing the slowly-rotating perturbation
are completely decoupled from the spin-$0$ monopole equations
governing the spherically symmetric background.
It follows that the behavior of inflation
in slowly-rotating black holes
is identical to that in spherical black holes.
This conclusion is true subject to the consistency condition
that the slowly-rotating perturbation remains small through inflation.
As seen in \S\ref{similaritymodel} below,
the slowly-rotating perturbation remains small,
so the solutions are indeed self-consistent.

The boundary conditions of the spherically symmetric self-similar solutions
are set at a sonic point outside the outer horizon,
where the infalling baryonic fluid accelerates from subsonic to supersonic.
The mass $\Mbh$ and charge $\Qbh$ of the black hole
can be defined to be the mass and charge that a distant observer
would see if there were no matter or charge outside the sonic point:
\begin{equation}
  \Mbh
  =
  M + {Q^2 \over 2 r}
  \ , \quad
  \Qbh
  =
  Q
  \quad
  \mbox{at the sonic point}
  \ .
\end{equation}
The extra mass $Q^2 / 2 r$ added to the interior mass $M$
is the mass-energy in the electric field outside the sonic point,
given no charge outside the sonic point.

For the spherically symmetric solutions,
there are three dimensionless boundary conditions at the sonic point:
(1) the mass accretion rate $\Mbhdot$;
(2) the charge-to-mass ratio $\Qbh/\Mbh$;
(3) the dark-matter-to-baryon density ratio $\rho_d / \rho_b$.
The solutions scale with the mass $\Mbh$ of the black hole.

\subsection{Similarity assumption}
\label{similarityassumption}

Self-similarity is the assumption
that the system possesses
conformal time translation invariance,
that is, there exists a conformal time $\tc$ such that
the system at any one conformal time
is a scaled copy of the system at any other conformal time.

The line-element~(\ref{perturbedlineelement})
can be cast in a form that has explicit conformal symmetry
by transforming the time and radial coordinates
$t$ and $r$
to conformal time and radial coordinates
$\tc$ and $\rc$ defined by
\begin{equation}
\label{conformalcoordinatetransformation}
  t
  =
  \tc
  \ , \quad
  \ln r
  =
  \rc
  +
  \tc
  \ .
\end{equation}
The fact that it is possible to choose the time coordinate
to be conformal time,
$t = \tc$,
reflects the gauge freedom in the choice of time coordinate.
In terms of the conformal coordinates $\tc$ and $\rc$,
the line-element~(\ref{perturbedlineelement}) is
\begin{eqnarray}
\label{selfsimilarlineelement}
  \dd s^2
  =&
  r^2 \Bigl\{
  - \,
  \left(
  \xi^t \dd \tc
  + \zeta^t
  \sin^2\!\theta
  \, \dd \phi
  \right)^2
  +
  \left(
  {\dd \rc \over \beta_r}
  +
  \xi^r \dd \tc
  +
  \zeta^r \sin^2\!\theta \, \dd \phi
  \right)^2
\nn
  &
  + \,
  \dd \theta^2
  +
  \left[
  \sin\theta
  \left( \dd \phi - \Psi \, \dd \rc \right)
  + \xi^\phi \, \dd \tc
  \right]^2
  \Bigr\}
  \ .
  \qquad
\end{eqnarray}
The circumferential radius $r$ (squared) appears as an overall conformal factor.
Conformal time translation symmetry requires that
all conformal inverse-vierbein coefficients,
the coefficients of differentials inside the braces
in equation~(\ref{selfsimilarlineelement}),
are
functions only of conformal radius $\rc$, not
of conformal time $\tc$.

The $\xi^m$ in the line-element~(\ref{selfsimilarlineelement})
are the tetrad-frame components of the Killing vector
associated with conformal time translation invariance,
\begin{equation}
\label{xim}
  {\partial \over \partial \tc}
  =
  e^m{}_\smalltc \, \partial_m
  \equiv
  r \xi^m \partial_m
  \ .
\end{equation}
In terms of the vierbein coefficients
of the line-element~(\ref{perturbedlineelement}),
the conformal Killing vector $\xi^m$ is
\begin{equation}
  \xi^m
  =
  \left\{
  {1 \over r \alpha} ,
  {1 \over \beta_r}
  \left(
  - \,
  {\beta_t \over r \alpha}
  +
  1
  \right) ,
  0 ,
  - (
  \Omega
  +
  \Psi
  ) \sin\theta
  \right\}
  \ .
\end{equation}
The scalar product of the conformal Killing vector $\xi^m$ with itself
defines the horizon function $\Delta$,
a coordinate and tetrad gauge-invariant scalar,
\begin{equation}
  \Delta
  \equiv
  - \xi_m \xi^m
  =
  \xi^t{}^2
  -
  \xi^r{}^2
  \ .
\end{equation}
The horizon function is positive outside the horizon,
zero at the horizon,
and negative inside the horizon,
reflecting the fact that
the Killing vector $\xi^m$ is respectively timelike, null, and spacelike
outside, at, and inside the horizon.
The scalar product of the conformal Killing vector $\xi^m$
with the radial 4-gradient $\beta_m$
follows from
$\partial \ln r / \partial \tc = 1$,
which implies
\begin{equation}
  \xi^m \beta_m
  =
  1
  \ .
\end{equation}

In self-similar solutions,
all quantities are proportional to some power of
circumferential radius $r$,
and that power can be determined by dimensional analysis.
Dimensional analysis shows that
the conformal coordinates
$\{ \tc, \rc, \theta, \phi \}$,
the coordinate metric $g_{\mu\nu}$,
and the vierbein perturbation
$\varphi_{mn}$
are all dimensionless.
In particular,
\begin{equation}
  \beta_m
  \ , \ 
  \xi^m
  \ , \ 
  \omega
  \ , \ 
  \psi
  \ , \ 
  \zeta^m
  \quad
  \mbox{are dimensionless}
  \ .
\end{equation}
The vierbein
$e_m{}^\mu$,
inverse vierbein
$e^m{}_\mu$,
tetrad-frame connections
$\Gamma_{kmn}$,
and
tetrad-frame Riemann tensor
$R_{klmn}$,
energy-momenta
$T_{mn}$,
and electromagnetic field
$F_{mn}$,
scale as
\begin{eqnarray}
  &
  e_m{}^\mu \propto r^{-1}
  \ , \quad
  e^m{}_\mu \propto r
  \ , \quad
  \Gamma_{kmn} \propto r^{-1}
  \ ,
\nn
  &
  R_{klmn} \propto r^{-2}
  \ , \quad
  T_{mn} \propto r^{-2}
  \ , \quad
  F_{mn} \propto r^{-1}
  \ .
\end{eqnarray}
Various specific quantities that appear in \S\ref{similarityequations} scale as
\begin{equation}
  W \propto r^0
  \ , \quad
  Q \propto r
  \ , \quad
  B \propto r
  \ , \quad
  a \propto r
  \ .
\end{equation}

\subsection{Similarity equations}
\label{similarityequations}

Four equations~(\ref{selfsimilarLbconservation})--(\ref{selfsimilarBeq})
govern the angular structure of
the slowly-rotating black hole,
one for each of the baryons, dark matter, gravity, and electromagnetic field.
For simplicity,
the equations in this subsection are for zero conductivity,
$\sigma = 0$,
as assumed in the model of \S\ref{similaritymodel}.
Equations for finite conductivity are given for reference in
Appendix~\ref{selfsimilareqsconducting}.

I follow
\cite{HP05a,HP05b}
in choosing
the integration variable $x$
to be the dimensionless baryonic time defined by
$\dd x \equiv r^{-1} \dd \tau$,
where
$\tau$ is the proper time along the worldline of the baryons.
Equivalently
\begin{equation}
  {\dd \over \dd x}
  =
  r \partial_t
\end{equation}
where
$\partial_t$
is the directed time derivative in the rest frame of the baryons.
The dimensionless baryonic time $x$ increases monotonically,
since proper time does.
The dark matter falls into the black along a different trajectory
than the baryons,
and it is therefore necessary to distinguish the dark matter radius $r_d$
and dark matter tetrad frame from the baryonic radius and tetrad frame,
as detailed in \cite{HP05b}.
I treat the baryonic frame as the default frame,
and for brevity drop the baryonic subscript $b$ from
$r$ and $\xi^m$
evaluated in the baryon frame.

In self-similar solutions,
the volume of a Lagrangian volume element
is proportional to
$\frac{4}{3} \pi r^3 \xi^r$.
Angular momenta $L$ of Lagrangian volume elements
can be defined to be
$L \equiv {\textstyle \frac{4}{3}} \pi r^3 \xi^r a ( \rho + p )$.
Explicitly, for each of the
baryon
(subscripted~$b$)
and dark matter
(subscripted~$d$)
streams,
whose equations of state are given by equations~(\ref{wb}) and (\ref{wd}),
\numparts
\begin{eqnarray}
  L_b
  \equiv
  {\textstyle \frac{4}{3}}
  \pi r^3 \xi^r a_b ( 1 + w ) \rho_b
  \ ,
\\
  L_d
  \equiv
  {\textstyle \frac{4}{3}}
  \pi r_d^3 \xi_d^r a_d \rho_d
  \ .
\end{eqnarray}
\endnumparts
Covariant conservation
$\zeta_n D_m ( T_e^{mn} + T_b^{mn} ) = 0$
of electromagnetic plus baryon energy
along the azimuthal Killing vector $\zeta^m$,
equations~(\ref{zetaDTe}) and (\ref{zetaDT}),
yields the following equation
of conservation of angular momentum of baryons
in their rest frame
[note that
for self-similar solutions
the tetrad-frame connection $h_r$ in eq.~(\ref{zetaDT}) is
$h_r = \partial_t \ln ( r \xi^r )$]:
\begin{equation}
\label{selfsimilarLbconservation}
  {Q B \over 6}
  +
  L_b
  =
  \mbox{constant}
  \ .
\end{equation}
Similarly, covariant conservation
$\zeta_n D_m T_d^{mn} = 0$
of dark matter energy
implies conservation of angular momentum of dark matter in its frame,
\begin{equation}
\label{selfsimilarLdconservation}
  L_d
  =
  \mbox{constant}
  \ .
\end{equation}
Equation~(\ref{selfsimilarLdconservation})
should be interpreted as meaning that the angular momentum $L_d$
is constant along the path of the dark matter,
not along the path of the baryons, which is different.
The dimensionless combination $L_d/r_d^2$
is however the same in any frame.

The operator
$\xi^m \partial_m$,
equation~(\ref{xim}),
vanishes
when acting on any dimensionless quantity.
Taking
$\xi^r$ times equation~(\ref{Einsteintogethert})
plus
$\xi^t$ times equation~(\ref{Einsteintogetherr})
yields an integral of motion
for the axial Weyl scalar $W$,
\begin{equation}
\label{selfsimilarW}
  W
  =
  {Q B \over 2 r^2}
  -
  {3 L_b \over r^2}
  -
  {3 L_d \over r_d^2}
  \ .
\end{equation}

Equation~(\ref{jphi})
governing the radial magnetic field scalar $B$,
coupled with equation~(\ref{zetaj})
and the assumption of vanishing conductivity,
yields a second order ordinary differential equation for $B$,
\begin{equation}
\label{selfsimilarBeq}
  - \,
  {1 \over 2 \Delta}
  \left[
  -1
  +
  \left(
  - \, {\Delta \over \xi^r} {\dd \over \dd x}
  +
  {\xi^t \over \xi^r}
  \right)^2
  \right]
  B
  +
  B
  +
  2 Q W
  =
  {a_b Q \over r \xi^r}
  \ .
\end{equation}

For self-similar solutions,
the vierbein coefficient $\omega$ is
(coordinate and tetrad) gauge-invariant,
and is related to the axial Weyl scalar $W$ by,
from equation~(\ref{W}),
\begin{equation}
\label{selfsimilaromega}
  - {1 \over 2 \xi^r} {\dd \omega \over \dd x}
  =
  W
  \ .
\end{equation}
The vierbein coefficients $\zeta^m$,
the tetrad-frame components
of the azimuthal Killing vector,
are not tetrad gauge-invariant.
A traditional choice of angular gauge
is the principal, or Weyl, gauge,
defined to be the frame in which the spin-$1$
components~(25)
of the Weyl tensor vanish.
In the principal tetrad,
\begin{equation}
\label{zetaprincipal}
  \zeta^m
  =
  - {a \Mbhdot \xi^m \over \Mbh}
  \ ,
\end{equation}
where the angular momentum parameter $a$ is
\begin{equation}
\label{aprincipal}
  a
  \equiv
  {\Mbh \over 6 \Mbhdot r^2 C \xi^r} {\dd W \over \dd x}
  \ .
\end{equation}
The factor of $\Mbhdot/\Mbh$
in equation~(\ref{zetaprincipal})
brings $a$ to the same scale as the angular momentum parameters
$a_b$ and $a_d$
of the baryons and dark matter,
equation~(\ref{a}).
The factor follows from scaling
from conformal time $\tc$ to Kerr-Newman time $t_{\rm KN}$
at the sonic point,
where
${\partial \ln r / \partial \tc} = 1$,
whereas
${\partial \ln r / \partial t_{\rm KN}} = \Mbhdot/\Mbh$,
so that
\begin{equation}
\label{tctkn}
  {\dd \tc \over \dd t_{\rm KN}}
  =
  {\Mbhdot \over \Mbh}
  \quad
  \mbox{at the sonic point}
  \ .
\end{equation}

The angular momentum parameter $\abh$ of the black hole itself
can be defined to be the value that would be measured by a distant observer
if there were no mass or charge outside the sonic point.
Matching the Kerr-Newmann~(\ref{knW})
and self-similar~(\ref{selfsimilarW})
values of the axial Weyl scalar $W$ at the sonic point yields
\begin{equation}
\label{abh}
  \abh \Mbh
  =
  {- \, r^2 W + Q B \over 3}
  =
  {Q B \over 6}
  +
  L_b
  +
  L_d
  \quad
  \mbox{at the sonic point}
  \ .
\end{equation}

\subsection{Similarity boundary conditions}

What angular boundary conditions must be adjoined to
the boundary conditions of the background spherically symmetric solution?

First,
the baryonic and dark matter angular momentum parameters $a_b$ and $a_d$,
equation~(\ref{a}),
must be constants at the sonic point,
independent of polar angle $\theta$.
This means that the angular flow patterns
of the infalling baryons and dark matter are dipoles,
with angular velocity proportional to $\sin\theta$.

The equation~(\ref{selfsimilarBeq}) governing the radial magnetic field scalar
$B$ is a second order differential equation,
so requires two boundary conditions.
However,
one boundary condition is fixed by the requirement that the solution
pass smoothly through the horizon,
which is a singular point of the differential equation.
I use an interative shooting method to locate the desired solution.
The remaining boundary condition on $B$
is in principle set by the requirement that $B$,
which is proportional to the radial component of a dipole magnetic field,
should go to zero at infinity (absent distant current sources).
However, the self-similar solutions generally do not continue
consistently to infinity.
Instead, I simply set $B$ equal to the Kerr-Newman value~(\ref{knB})
at the sonic point.

The angular momentum parameters $a_b$ and $a_d$ of the baryons and dark matter,
together with the value of $B$ at the sonic point,
determine the total angular momentum parameter $\abh$ of the black hole
through equation~(\ref{abh}).
Since all angular quantities scale in proportion to $\abh$,
the angular behavior of the solutions is determined in effect
by a single dimensionless boundary condition,
the ratio $a_d/a_b$ of the dark matter to baryonic angular momentum parameters
($B$ being fixed to its Kerr-Newman value at the sonic point).


\slowkerrvarsfig

\subsection{Similarity model}
\label{similaritymodel}

Figure~\ref{slowkerrvars}
shows a self-similar solution
with accretion rate $\Mbhdot = 0.01$,
charge-to-mass $\Qbh/\Mbh = 0.8$,
and dark-matter-to-baryon density ratio
$\rho_d / \rho_b = 0.1$
at the sonic point.
This is the same model shown in Figs.~4, 6, and 7 of \cite{HA08}.
The ratio $a_d/a_b$ of dark matter to baryon angular momentum parameters
has been set to $1$ far from the black hole
\begin{equation}
  {a_d \over a_b}
  =
  1
  \ .
\end{equation}

The accretion rate $\Mbhdot = 0.01$
(which essentially means that the sonic point is expanding at
$0.01$ of the speed of light, as measured by a distant observer)
is much larger than any typical astronomical accretion rate.
The advantage of a large accretion rate is that
it is possible to follow the complete
evolution of the spacetime through inflation to collapse
without quantities overflowing numerically.
The smaller the accretion rate,
the more rapidly inflation exponentiates \cite{HA08},
triggering numerical overflow.

The top panel of
Figure~\ref{slowkerrvars}
shows the various components of
the tetrad-frame energy-momentum tensor
measured in a frame that is center-of-mass in the radial direction,
$T^{tr} = 0$,
and principal in the angular direction,
where the spin-$1$ components of the Weyl tensor,
equation~(25),
vanish.
The center-of-mass energy density and pressure
$T^{tt}$ and $T^{rr}$
inflate rapidly at a radius $r \approx 0.4$
close to where the inner horizon would have been in the absence of inflation.
The center-of-mass energy and pressure
arise almost entirely from the streaming energy and pressure
produced by the relativistic counter-streaming between the
outgoing baryons and the ingoing dark matter.
The proper energy and pressure of either of the baryon or dark matter
streams in their own frames change only modestly.

The angular components
$T^{t\phi}$ and $T^{r\phi}$
in the top panel of Figure~\ref{slowkerrvars}
are plotted in units such that
the black hole angular parameter is unity, $\abh = 1$,
and should be multiplied by the true $\abh$,
a small number.
Thus although the Figure appears to show that,
in the Kerr-Newman regime $r \gtrsim 0.4$,
the angular components
$T^{t\phi}$ and $T^{r\phi}$
are comparable to the radial components
$T^{tt}$ and $T^{rr}$,
in fact the angular components are much smaller.
In the inflationary regime $r \lesssim 0.4$,
the angular components
$T^{t\phi}$ and $T^{r\phi}$
inflate,
but they inflate more slowly than the radial components
$T^{tt}$ and $T^{rr}$.
Thus the angular components,
already a small perturbation in the Kerr-Newman regime,
become an even smaller perturbation in the inflationary regime.

The upper middle panel of
Figure~\ref{slowkerrvars}
shows the axial Weyl and electromagnetic scalars $W$ and $B$,
equations~(\ref{W}) and (\ref{Br}),
while the lower middle panel
shows the gauge-invariant vierbein perturbation $\omega$,
equation~(\ref{omega}).
All three quantities remain well-behaved during inflation,
showing no sign of any inflationary behavior.
This demonstrates that the assumption of a slowly-rotating perturbation
is consistent:
the perturbation remains small throughout inflation.

Note that the $\omega$ obtained from equation~(\ref{selfsimilaromega})
is a conformal angular velocity.
Figure~\ref{slowkerrvars}
shows $\omega$ multiplied by $\Mbhdot/\Mbh$,
equation~(\ref{tctkn}),
which converts $\omega$ to a true angular velocity
that can be compared directly to the
Kerr-Newman angular velocity~(\ref{knomega}).

The lower panel of
Figure~\ref{slowkerrvars}
shows various angular momentum parameters $a$.
The dark matter angular momentum parameter $a_d$ is constant,
reflecting the fact that the dark matter is neutral and pressureless,
and therefore falls freely with constant angular momentum.
The baryonic angular momentum parameter $a_b$ decreases inwards,
as the charged baryons lose angular momentum into the magnetic field,
in accordance with equation~(\ref{selfsimilarLbconservation}).
The parameter $a$, equation~(\ref{aprincipal}),
is defined to be the angular momentum parameter in the principal frame.
In the Kerr-Newman geometry, this angular momentum parameter would be constant,
and equal to $1$ in the units $\abh = 1$ used in the Figure.
Figure~\ref{slowkerrvars}
shows that $a$ in the self-similar solution is indeed close to $1$
in the Kerr-Newman regime $r \gtrsim 0.4$.
What appears to be a glitch in $a$ at $r \approx 0.63$
arises because $\dd W / \dd x$ and $C$,
whose ratio yields $a$, equation~(\ref{aprincipal}),
go through zero at very slightly different radii
(in Kerr-Newman, the zeros occur at exactly the same radius).
No physical divergence is associated with the glitch.
All three angular momentum parameters remain well-behaved
in the inflationary regime: they do not inflate.

\Brstarfig

The angular structure of the self-similar solution
is determined by four equations,
of which three are algebraic,
equations~(\ref{selfsimilarLbconservation})--(\ref{selfsimilarW}),
and the fourth is a second order differential equation for
the radial magnetic field scalar $B$,
equation~(\ref{selfsimilarBeq}).
Only the fourth equation poses a challenge to understanding its behavior.
Figure~\ref{Brstar} shows that
in the Kerr-Newman regime
$r \gtrsim 0.4$,
the magnetic $B$ field is closely approximated by the
Kerr-Newman form~(\ref{knB}).
Inflation takes place just above the inner horizon,
and in the inflationary regime
$r \lesssim 0.4$,
the horizon function $\Delta$ is tiny,
and $\xi^t / \xi^r \approx - 1$.
In this regime the differential
equation~(\ref{selfsimilarBeq}) for $B$
simplifies to
\begin{equation}
\label{inflationaryBeq}
  \left[
  -1
  +
  \left(
  - \, {\Delta \over \xi^r} {\dd \over \dd x}
  -
  1
  \right)^2
  \right]
  B
  =
  0
  \ .
\end{equation}
Introduce the Regge-Wheeler coordinate $r^\ast$ defined by
\begin{equation}
  {\dd \over \dd r^\ast}
  \equiv
  - \,
  {\Delta \over \xi^r} {\dd \over \dd x}
  \ .
\end{equation}
Then equation~(\ref{inflationaryBeq}) simplifies to
\begin{equation}
  {\dd \over \dd r^\ast}
  \left(
  {\dd \over \dd r^\ast} - 2
  \right)
  B
  =
  0
  \ ,
\end{equation}
whose general solution is
\begin{equation}
\label{Binflationapproximation}
  B =
  B_0
  +
  B_1 \ee^{2 r^\ast}
\end{equation}
where $B_0$ and $B_1$ are constants.
Figure~\ref{Brstar} shows
a fit to the approximation~(\ref{Binflationapproximation})
with the constants $B_0$ and $B_1$ treated as free parameters,
and it is seen that the fit is excellent.
Analytic and numerical investigation
indicates that the constant $B_1$ depends
on boundary conditions (such as $\rho_d/\rho_b$ or $a_d/a_b$),
and does not tend to some universal limit
in the limit of small accretion rates
(for example, $B_1$ does not vanish).

\section{Small accretion rate}
\label{smallaccretion}

The self-similar model of the previous section,
\S\ref{similarity},
has the virtue of yielding exact solutions,
but it is unrealistic in the sense that it requires
a finely-tuned accretion flow,
in which the accreting streams have
a small dipole angular velocity
that remains steady over the age of the black hole.

In fact however,
as will now be shown,
for
sufficiently
small accretion rates,
condition~(\ref{smallaccretioncondition}),
the instantaneous angular motion of the accretion flow
has a negligible effect on the spacetime geometry
of the slowly rotating black hole.
For small enough accretion rates,
the equations describing the angular spacetime structure
of a slowly-rotating black hole simplify to
equations~(\ref{Wsmallaccretion})
and (\ref{Bsmallaccretion}),
independent of the instantaneous angular motion of the accreting streams.

For small accretion rates,
the interior charge $Q$ can be treated as effectively constant
as a function of radius,
\begin{equation}
\label{Qconstant}
  Q
  \approx
  \Qbh
  =
  \mbox{constant}
  \ .
\end{equation}
Initially, before inflation, the charge is constant
simply because the small accretion rate implies a small charge density.
During inflation,
as first pointed out by
\cite{Ori91},
Lagrangian volume elements
remain little distorted
even though the tidal force (Weyl curvature)
exponentiates hugely.
The physical reason for the mild distortion is that
a volume element experiences only a tiny proper time
during inflation,
so that the tidal force, though huge,
has too little time to do damage.
Because volume elements change little during inflation,
and because charge density is conserved and the initial charge density is small,
therefore the charge density remains always small,
and the interior charge $Q$ is effectively constant.

For constant interior charge $Q$,
the Einstein equation~(\ref{Einsteintogetherr})
integrates along the worldline of any tetrad frame to
\begin{equation}
\label{Einsteintogetherrintegrated}
  r^2 W
  =
  \Qbh B
  -
  3 \abh \Mbh
  +
  8\pi
  \sum_{{\rm streams}}
  \int_{{\rm sonic}}
  \!\!\!\!\!\!
  a
  ( \rho + p ) u^r
  r^2
  {\dd r \over \beta_t}
  \ .
\end{equation}
Here the integral with respect to proper time $\tau$
along the worldline of the tetrad
has been converted to an integral with respect to radius $r$
using $\dd r / \dd \tau \equiv \partial_t r = \beta_t$.
The constant
$- 3 \abh \Mbh$
of integration
in the integral~(\ref{Einsteintogetherrintegrated})
is established at the sonic point,
where equation~(\ref{abh}) holds.

For small accretion rates,
the integral over either of the baryonic or dark matter streams
in equation~(\ref{Einsteintogetherrintegrated})
is small
\begin{equation}
\label{streamintegral}
  \int
  a
  ( \rho + p ) u^r
  r^2
  {\dd r \over \beta_t}
  \approx
  0
  \ .
\end{equation}
Before inflation, the integral is small
because the small accretion rate
implies a small proper density and pressure $\rho$ and $p$.
During inflation,
the proper density and pressure $\rho$ and $p$
of a stream change only mildly
(because Lagrangian volume elements distort mildly),
and,
thanks to conservation of angular momentum,
the angular momentum parameter $a$ of the stream
likewise changes little.
The only factor that could potentially become large
in the integrand of the integral~(\ref{streamintegral})
is the ratio
$u^r/\beta_t$.
The integrals in equation~(\ref{Einsteintogetherrintegrated})
can be evaluated in any tetrad frame
(the same frame for both streams);
one option is to choose the no-going frame,
defined to be the frame where $\beta_r =0$.
Relative to the no-going frame,
the streaming velocity $u^r$ and $\beta_t$
both increase exponentially,
but they increase approximately in proportion to each other
\cite{HA08}.
The result is that
the integral~(\ref{streamintegral})
remains small because $\rho + p$ is small,
and the other factors in the integrand do not spoil that conclusion.

The integral over either stream
in equation~(\ref{Einsteintogetherrintegrated})
involves the product
$a ( \rho + p )$
of the angular momentum parameter
and the proper density plus pressure.
The integrals are small compared to the other terms
in equation~(\ref{Einsteintogetherrintegrated})
provided that $a ( \rho + p ) \ll \abh$
in geometric units ($c = G = \Mbh = 1$).
If the streams are slowly rotating, $a \sim \abh$,
then the integrals are small provided that
$\rho + p \ll 1$.
But if the accretion rate is sufficiently small
that it satisfies the stronger condition that
\begin{equation}
\label{smallaccretioncondition}
  \rho + p \ll \abh
  \quad
  (\mbox{geometric units})
  \ ,
\end{equation}
then the integrals remain small even if the angular momentum parameter $a$
is large, of order unity in geometric units.
The fact that the integrals remain small,
equation~(\ref{streamintegral}),
even for large angular momenta means
that the instantaneous angular motion of the accreting streams
has effectively negligble effect on the spacetime structure
of the slowly-rotating black hole,
even in the case that
the accreting streams have large angular momentum
and any angular motion,
not just cylindrically symmetric rotation about the rotation axis.
Of course, ultimately the angular momentum of the black hole
is equal to the cumulative angular momentum of the accreted streams;
but the instantaneous angular motion is essentially irrelevant.

In the approximations~(\ref{Qconstant}) and (\ref{streamintegral}),
valid for small enough accretion rates,
equation~(\ref{Einsteintogetherrintegrated}) simplifies to
\begin{equation}
\label{Wsmallaccretion}
  r^2 W
  \approx
  \Qbh B
  -
  3 \abh \Mbh
  \ ,
\end{equation}
which relates the axial Weyl scalar $W$
to the axial magnetic field scalar $B$
without any dependence on the baryon or dark matter streams.

To complete the equations governing the spacetime structure
for small accretion rates,
equation~(\ref{Wsmallaccretion})
must be supplemented by
an equation for the magnetic field $B$,
specifically equation~(\ref{jphi})
with, since the accretion rate is small,
vanishing right hand side.
Self-similar solutions provide a guide to what happens.
As illustrated in Figure~\ref{Brstar},
$B$ is well-approximated by the slowly-rotating
Kerr-Newman solution~(\ref{knB})
before inflation sets in,
and by approximation~(\ref{Binflationapproximation})
in the inflationary regime,
The approximation~(\ref{Binflationapproximation})
depends on the Regge-Wheeler coordinate $r^\ast$,
which has the property that,
as illustrated in Figure~\ref{Brstar},
the derivative $\dd r^\ast / \dd \ln r$
goes through a maximum near the onset of inflation,
and then becomes progressively smaller during and after inflation.
The net effect is that
$B$
flattens out and becomes constant during and after inflation,
which is true even though the constant $B_1$
in the approximation~(\ref{Binflationapproximation}),
which is effectively determined by matching to the Kerr-Newman
approximation~(\ref{knB}) at the onset of inflation,
is non-zero.
For small accretion rates,
the transition from the Kerr-Newman regime
to the regime where $B$ is constant
becomes increasingly abrupt,
and $B$ simplifies to
\begin{equation}
\label{Bsmallaccretion}
  B
  \approx
  {2 \abh \Qbh \over \max( r , r_- )}
  \ ,
\end{equation}
where $r_- = \Mbh - \sqrt{\Mbh^2 - \Qbh^2}$
is the radius of the erstwhile inner horizon,
which is destroyed by inflation.
The approximation~(\ref{Bsmallaccretion})
has been derived on the assumption of self-similarity,
which is premised on the assumption of constant accretion rate $\Mbhdot$.
However,
since the approximation~(\ref{Bsmallaccretion})
is independent of the value of the accretion rate,
it may be guessed that the approximation~(\ref{Bsmallaccretion})
remains valid even for a variable accretion rate,
just so long as the accretion rate remains small.

The axial spin-$0$ Weyl scalar $W$ is in a fundamental sense
the (coordinate and tetrad) gauge-invariant quantity
that describes the angular part of the gravitational field.
The Weyl scalar $W$ is related to
the angular vierbein coefficients $\omega$ and $\psi$
by equation~(\ref{W}),
which in the self-similar case becomes equation~(\ref{selfsimilaromega}),
which depends only on $\omega$.
Converted to a derivative with respect to $\ln r$ instead of $x$,
the self-similar equation~(\ref{selfsimilaromega})
for $\omega$ becomes
$- ( \beta_t / 2 \xi^r ) {\dd \omega / \dd \ln r} = W$.
In the inflationary regime,
the coefficient $- \beta_t / 2 \xi^r$ is so large,
for small accretion rates,
that $\dd \omega / \dd \ln r$ is tiny,
so that $\omega$ is essentially constant as a function of radius $r$
(beware: this statement fails deep into the
post-inflationary collapse regime).
The result is that, for small accretion rates,
the vierbein coefficient $\omega$ is
approximately equal to the Kerr-Newman form~(\ref{knomega})
prior to inflation,
and to a constant during and after inflation:
\begin{equation}
\label{omegasmallaccretion}
  \omega
  \approx
  {\abh \over \max(r,r_-)^3}
  \left(
  2 \Mbh - {\Qbh^2 \over \max(r,r_-)}
  \right)
  \ .
\end{equation}
As with equation~(\ref{Bsmallaccretion}),
equation~(\ref{omegasmallaccretion})
has been derived on the assumption of self-similarity,
but in fact the equation
is independent of the value of the accretion rate,
so it may be guessed that the approximation~(\ref{omegasmallaccretion})
remains valid even for an accretion rate that varies with time,
just so long as the accretion rate remains small.

Equations~(\ref{Wsmallaccretion})--(\ref{omegasmallaccretion})
look enticingly simple,
but one should be wary of interpreting them too literally,
especially equation~(\ref{omegasmallaccretion}).
For example, if the approximation $\omega \approx \mbox{constant}$
in the inflationary regime were taken literally,
then it would imply,
from equation~(\ref{selfsimilaromega}),
that $W \approx 0$,
which is false.
Thus it would be incorrect to insert the
approximation~(\ref{omegasmallaccretion})
into the line-element~(\ref{perturbedlineelement})
and imagine that the consequent Riemann tensor would be valid.
The problem is that inflation generates exponentially huge numbers,
and something that by itself appears sensibly equal to zero
may become significant when multiplied by a huge number.

\section{Limitation}
\label{limitation}

The slowly-rotating models considered in this paper
have the limitation that the rates of accretion of ingoing and outgoing
streams are assumed to be constant as a function of angular position
about the black hole.
This limitation is inherited from the assumption
that the slow rotation is a small perturbation
of a background spacetime that is spherically symmetric.
The previous section
showed that the angular motion of the accreting streams
has negligible effect, for small accretion rates.
But the angular motion is distinct
from the angular dependence of the accretion rate.

It would be possible to consider,
in addition to the slowly rotating perturbation
considered in this paper,
perturbations to the accretion rate as a function of angular position
over the black hole.
This could be the subject of future investigation.

\section{Summary}

The internal structure of a slowly rotating black hole
that is undergoing inflation at (just above) its inner horizon
has been derived.
The approach has been to introduce a slowly-rotating
perturbation to a spherically symmetric black hole undergoing inflation.
It has been shown that the rotational perturbation remains small
throughout inflation, so the perturbation assumption is self-consistent.

The equations governing the angular behavior, \S\ref{slowlyrotating},
decouple from the radial behavior.
Consequently all of the conclusions concerning inflation
in spherically symmetric black holes
carry through unchanged for slowly rotating black holes.
This conclusion supports
\cite{PI90}'s
conjecture that
inflation in rotating black holes is probably similar to
inflation in spherical black holes.

Exact self-similar solutions for slowly rotating black holes have been obtained,
\S\ref{similarity}.
The self-similar solutions require a special accretion flow,
in which the accreting streams have a small dipole angular velocity
that remains steady over the age of the black hole.

It has been shown, \S\ref{smallaccretion},
that for
sufficiently small accretion rates
the instantaneous angular motion of the accretion flow
has a negligible effect on the spacetime geometry of
the slowly rotating black hole.
The angular structure of the black hole is described
by the axial spin-$0$ Weyl and magnetic field scalars $W$ and $B$,
which for small accretion rates are given approximately by
equations~(\ref{Wsmallaccretion}) and (\ref{Bsmallaccretion}),
independent of the instantaneous angular motion of the accretion flow.

The slowly-rotating solutions obtained in this paper
all have an accretion rate that is constant as a function of angular position
over the black hole,
a limitation inherited from the assumption that the slow rotation
is a small perturbation from spherical symmetry.

\ack
This work was supported in part by NSF award
AST-0708607.

\appendix

\section{Finite conductivity}
\label{selfsimilareqsconducting}

This Appendix gives, for reference,
the self-similar equations for the case of non-vanishing
baryonic electrical conductivity $\sigma$.
Equation~(\ref{selfsimilarLbconservationconducting})
replaces equation~(\ref{selfsimilarLbconservation}),
while equation~(\ref{selfsimilarBeqconducting})
replaces equation~(\ref{selfsimilarBeq}).

To admit self-similar behavior,
the conductivity must have dimension
$\sigma \propto r^{-1}$.
If the conductivity is assumed to depend only on the baryon
proper density $\rho_b$,
then, since $\rho_b \propto r^{-2}$, it follows that
[the following repeats eq.~(48) of \cite{HP05a}]
\begin{equation}
  \sigma
  =
  {\kappa \rho^{1/2} \over ( 4 \pi )^{1/2}}
  \ ,
\end{equation}
where $\kappa$ is a phenomenological dimensionless
conductivity coefficient.
A dimensionless conductivity $s$ can be defined as
[the following repeats eq.~(51) of \cite{HP05a}]
\begin{equation}
  s
  \equiv
  4 \pi \sigma r
  =
  \kappa
  ( 4 \pi r^2 \rho )^{1/2}
  \ .
\end{equation}

In terms of the dimensionless electrical conductivity $s$,
the equation expressing conservation of
the angular momentum of baryons is
\begin{equation}
\label{selfsimilarLbconservationconducting}
  {\dd \over \dd x}
  \left(
  {Q B \over 6}
  +
  L
  \right)
  =
  -
  {s Q B \over 3}
  \ ,
\end{equation}
and the equation governing the radial magnetic field scalar $B$ is
\begin{equation}
\label{selfsimilarBeqconducting}
\fl
  - \,
  {1 \over 2 \Delta}
  \left[
  -1
  +
  \left(
  - \, {\Delta \over \xi^r} {\dd \over \dd x}
  +
  {\xi^t \over \xi^r}
  \right)^2
  \right]
  B
  +
  B
  +
  2 Q W
  =
  {l Q \over r \xi^r}
  ( 1 + s \xi^t )
  -
  {s \over 2} {\dd B \over \dd x}
  \ .
\end{equation}

\end{document}